\documentclass[amssymb, nofootinbib, superscriptaddress]{aipproc} 
\layoutstyle{6x9}
\usepackage{amsmath}
\usepackage{amsfonts}
\usepackage{graphicx}
\usepackage{psfrag}
\usepackage{array}
\usepackage{bbm}
\usepackage{hyperref}

\newcommand{\ie}{\textit{i.e.}~}

\newcommand{\f}{\frac}

\newcommand{\R}{\mathbb{R}}
\newcommand{\C}{\mathbb{C}}

\newcommand{\cc}{{\cal D}}

\newcommand{\be}{\begin{equation}}
\newcommand{\ee}{\end{equation}}
\newcommand{\bes}{\begin{eqnarray}}
\newcommand{\ees}{\end{eqnarray}}
\newcommand{\mone}{^{-1}}

\def\dia{\diamond}

\def\f{\frac}  

\def\tl{\widetilde}

\def\ka{\kappa}
\def\cc{{\cal C}}

\def\act{\rhd}
\def\mn{{\mu\nu}}

\def\ie{{i.e. \/}}

\def\dr{{\rightarrow}}

  \def\cc{{\cal C}}    \def\ggg{{\cal G}}
\def\hh{{\cal H}}     \def\mm{{\cal M}} \def\nn{{\nonumber}}
 \def\ppp{{\cal P}}

\def\hphi{{\hat \phi}}

\def\act{{\, \triangleright\, }}

\def\rltimes{{\,\triangleright\hspace{-1.45mm}< \,}}

\def\double{{\triangleright\hspace{-0.5mm}\triangleleft}}

\newcommand{\su}{\mathfrak{su}}
\newcommand{\an}{\mathfrak{an}}
\newcommand{\so}{\mathfrak{so}}

\newcommand{\AN}{\mathrm{AN}}

\newcommand{\SO}{\mathrm{SO}}
\newcommand{\SL}{\mathrm{SL}}

\begin{document}

\title{Field theories  with  homogenous momentum space }

\author{Florian Girelli}{
  address={School of Physics A28, Sydney University, 2006 Sydney, Australia},
  email= girelli@physics.usyd.edu.au
}
\author{Etera Livine}{
  address={Laboratoire de Physique, ENS Lyon, CNRS UMR 5672, 46 All\'ee d'Italie, 69007 Lyon, France},
  email= etera.livine@ens-lyon.fr
}

\begin{abstract}
We discuss the construction of a scalar field theory with momentum space given by a coset. By introducing a generalized Fourier transform, we show how the dual scalar field theory actually lives in Snyder's space-time. As a side-product we identify  a star product realization of Snyder's non-commutative space, but also the deformation of the Poincar\'e symmetries necessary to have these symmetries realized in Snyder's space-time. A key feature of the construction is that the star product is non-associative.   \end{abstract}

\maketitle

\section{Introduction}
In the recent years, there has been a growing interest regarding the non-commutative spaces of the Lie algebra type, \ie spaces where  coordinates can be represented as Lie algebra generators. These types of models have indeed appeared in the context of quantum gravity,  as candidates for the flat semi-classical limit. In 3d, it has been actually shown that a $\su(2)$ space-time correctly describes the flat semi-classical limit of Euclidian 3d quantum gravity \cite{etera-laurent}. Meanwhile, in 4d, different arguments have been proposed to argue that $\an_3$, \ie $\ka$-Minkowski, is the right flat semi-classical limit \cite{dsr}.
These non-commutative spaces are flat spaces, that is it is possible to introduce a deformation of the Poincar\'e group, in particular of the translations, such that the non-commutativity is  consistent with the Poincar\'e transformations (for a classification of the Poincar\'e group deformations see for example \cite{zak}).  If the space-time is given in terms of a Lie algebra, momentum space (or \emph{Fourier space}) is in turn defined as a (non-abelian) Lie group and acquires some curvature.
Under the Fourier transform from the Lie algebra coordinate space to the Lie group momentum space, a scalar field theory defined over a non-commutative space-time of the Lie algebra type can therefore be seen as a scalar field theory defined on a \emph{group manifold}.

\medskip

This is a well-known theory in the context of quantum gravity, where one calls it a \emph{group field theory} (GFT). These GFT are  used to generate spinfoam models \cite{GFT}.  The fact that group field theories are at the same time a fundamental object in the spinfoam framework and a natural object to consider in non-commutative geometry framework is very interesting and certainly a sign of fruitful future interplays between these two different frameworks. Recently this connection was used to relate in a new and original way the different flat semi-classical  space-time  to  some spinfoam models \cite{dsr-GFT}.

\medskip

In general momentum space can be seen as an homogenous space. For example in the usual case, one considers the Poincar\'e group $\ppp \sim \SO(n-1,1)\rltimes \R^n$, where the Lorentz group acts on $\R^n$ on the left. Momentum space can then be obtained as $\mm\sim \ppp/\SO(n-1,1)\sim \R^n$. The most natural generalization of this construction is to consider a pair of Lie groups $G,\, H$ which acts on each other in a consistent way to obtain a "double cross product group" $G\double H$ \cite{majid}. A typical way to have such pair is to consider a group $\ggg$ which factorizes  $\ggg\sim G\cdot H$.  Momentum space is then $\mm\sim \ggg/H\sim G$. In this case the scalar field will be a representation of the quantum group introduced by Majid, the bicrossproduct quantum group \cite{majid}. The most well known example is given by the Iwasawa decomposition of $\SO(4,1)$ \cite{klymyk}, which encodes that $\SO(4,1)$ is factorizable  $\SO(4,1)\sim \AN_3 \cdot \SO(3,1)$. Momentum space is then obtained as the homogenous space $\mm\sim\AN_3\sim \SO(4,1)/\SO(3,1)$, which has a group structure. The corresponding  Lie algebra is $\an_3$ and defines the $\ka$-Minkowski space-time. We shall present in further details the use of the bicrossproduct construction in the context of GFT in \cite{bicross en prep}.

\medskip

An homogenous space  does not have  in general  a group structure. By considering  the Cartan decomposition of a group $G$ using the maximal subgroup $H$, the homogenous space $G/H$ is only a \emph{coset}.   These types of momentum spaces are again interesting both from the non-commutative geometry  and the spinfoam GFT  points of view. Indeed, in 1947 Snyder introduced one of the first examples of non-commutative geometry \cite{snyder} which  can be encoded in terms of coordinates    as
$$X_\mu\sim\frac{1}{\kappa}J_{p\mu} \in \so(p-1,1)/\so(p-2,1), \qquad [X_\mu,X_\nu]=i\frac{1}{\ka^2}J_{\mn},$$
where $J_\mn$ is an infinitesimal Lorentz transformation and $\ka$ is the Planck scale. Momentum space is  the de Sitter space $dS$ seen as the homogenous space obtained from the Cartan decomposition $dS\sim \SO(4,1)/\SO(3,1)$. There has been up to now very few attempts to identify the deformation of the Poincar\'e group associated to this non-commutative space and to construct a scalar field theory transforming under these deformed symmetries. On the other hand the GFT which generates spinfoam models for 4d quantum gravity are typically field theories defined on the product of coset spaces $G/H$, where for example $G=\SO(4)$ and $H=\SO(3)$ in the Euclidian case \cite{gft-4d}.

\medskip

We intend here to construct a scalar field theory defined on a coset seen as momentum space, and introduce a generalized Fourier transform to analyze the properties of the dual space-time (for different approaches regarding the definition of a scalar field theory in Snyder space-time see \cite{othersnyder}).

\medskip

We show first how using the coset structure we can define  a momenta addition and a convolution product (both of them will be non-associative). These are the necessary ingredients to construct the scalar field action. We show then how one can introduce a generalized Fourier transform which will allow to define a non-commutative star product as the dual of the convolution product.  We pinpoint that this star product is in fact a realization of Snyder's non-commutative space-time. Before ending with some concluding remarks, we discuss the notion of symmetries in this non-commutative space.

\section{Momentum space as a coset }

In this section we  construct an action for a scalar field theory defined on the right coset $G/H$. In the following we work with the 3d hyperboloid $\hh_3=\SO(3,1)/\SO(3)=G/H$ as a guiding example, but all the steps described here can be extended to more general cosets. Any element $g\in G$ can be written as $g=ah$, with $a\in G/H$ and $h\in H$. We note $[da]$  the measure over the coset  and  $[dh]$ the Haar measure  on $H$.

Momentum is identified as the coordinate system on the coset space.  The 3d hyperboloid is defined as
\be\label{hyperboloid}
\hh_3\sim \SO(3,1)/\SO(3)\sim \{v_\mu\in\R^4, \, v_0^2-v_i^2=1\}.
\ee
We define the Snyder coordinates  $P_i= \ka \frac{v_i}{v_0}$ as parameterizing our 3-momentum space~\footnotemark. The $\SO(3,1)$-invariant measure on is $\hh_3$:
$$
[dP]
\,=\,
\ka^3\,d^4v \,\delta(v_0^2-v_i^2-1)
\,=\,
d^3P\,\left(1- \frac{\vec P^2}{\kappa^2}\right)\mone.
$$
\footnotetext{{Of course different choices of parametrization are possible. For example, using $p_i= \ka v_i$ \cite{snyder gili}, we would get a different deformed momentum addition:
$$
\vec p_1 \oplus \vec p_2\,=\,
\left(\gamma_2+ \frac{1}{1+\gamma_1}\frac{\vec p_1\cdot \vec p_2}{\ka^2}\right)\vec p_1 +\,\vec p_2.
$$
The main difference between $\vec p$ and the Snyder momentum $\vec P$ is that $P$ is bounded by $\ka$ while $p$ remains unbounded.}}
A coset element is given by a  boost $a=e^{i \f\eta2 \vec b\cdot \vec K}\,\in\SO(3,1)$, where $K_i\equiv J_{0i}$ are the boost generators,  $\eta$ the angle of the boost  and the unitary vector $\vec b$ its direction. For the explicit calculations, we actually work in the spinorial representation of $\SO(3,1)$, given in term of $2\times 2$ group elements belonging to $\SL(2,\C)$. Such a group element acts on the 3+1d-Minkowski space and boosts the unit time-like vector $(1,0,0,0)$ to a vector $v_\mu$:
$$
v\,=\, a\rhd\,(1,0,0,0)\,=\,(\cosh \eta,\sinh\eta\,\vec{b}).
$$
This defines a Snyder momentum expressed in terms of $\eta$ and $\vec b$:
\be \label{snyder coordinates} \vec P= \frac{\ka}{v_0}\,\vec{v}\,=\, \ka\tanh\eta\, \vec b.\ee
From this definition of the section, one can check that the measure $[da]$ on the coset is as expected $[da]=[dP]$.
The group product on $\SO(3,1)$ induces a product on the coset between elements $a_i\in G/H$, which we note $a_{12}\equiv a_1\cdot a_2$. In turn, this product induces a momenta addition:
\be\label{notation product non assoc}
a_1a_2 =a_{12}h_{12}\, \dr \, a_{12}=e^{i \f{\eta_1} 2\vec b_1\cdot \vec K} \cdot e^{i\f{\eta_2} 2\vec b_2\cdot \vec K}=e^{i \f{\eta_{12}}2(\vec b_1\oplus \vec b_2)\cdot \vec K}   \quad a_i\in G/H, \, h_{12}\in H.  \ee
The $\SO(3)$-group element $h_{12}$ is uniquely defined and an explicit formula can be found in \cite{SR}. The relation \eqref{snyder coordinates} is then   used together with \eqref{notation product non assoc} to define the addition in terms of the Snyder coordinates~:
\be\label{non assoc add}
\vec P_1 \oplus \vec P_2= \frac{1}{1+\frac{\vec P_1\cdot \vec P_2}{\ka^2}}\left(\left(1+ \frac{\gamma_1}{1+\gamma_1}\frac{\vec P_1\cdot \vec P_2}{\ka^2}\right)\vec P_1 +  \frac{1}{\gamma_1}\vec P_2\right), \quad \gamma_1= \frac{1}{\sqrt{1-\frac{\vec P^2_1}{\ka^2}}},
\ee
This addition is non-associative due to the coset structure and therefore the ordering in which we group the addition is important.

We would like to emphasize here the parallel with Special Relativity. Indeed in this context the hyperboloid $\hh_3$ is the space of 3d speeds $\vec \upsilon$. The embedding space $\R^4$ is the space of relativistic speeds $v_\mu$. The usual choice of coordinates on $\hh_3$ is given by $\vec \upsilon = c\frac{\vec v}{v_0}= c\tanh\eta \vec b$. By noting $v_0=\gamma$, we have equivalently $\vec v= \gamma \vec \upsilon/c$, the standard expression for the relativistic speed. Finally the addition of speeds  is generated by the coset product  in \eqref{notation product non assoc}, and one obtains explicitly this addition by considering \eqref{non assoc add} and replacing $\vec P_i$ by $\vec \upsilon_i$ and $\ka$ by $c$ \cite{SR}.

To build an action, we consider the distributions $\phi(a)$ and $\delta(a)$ which are respectively a function (the real scalar field) on the coset seen as a distribution and the Dirac delta function. The $\delta$-distribution on the coset is inherited from the group structure:
$$
\delta_{\hh_3}(a)=\int [dh]\,\delta_{\SO(3,1)}(ah)=\delta^{(3)}(\vec{P}).
$$
In the following, we will drop the indices $\hh_3$ or $\SO(3,1)$.
We introduce the convolution product $\dia$ using the product on the coset \eqref{notation product non assoc}.
\bes\label{coset convolution}
\phi \dia \psi (a)&=& \int [da_1][da_2][dh]\, \phi(a_1)\psi(a_2)\delta(a\mone a_1a_2h)\nn\\
&=& \int [da_1][da_2]\, \phi(a_1)\psi(a_2)\delta(a\mone \cdot a_{12}).
\ees
We insist on the fact  that the $\delta$-distribution in the first line is on the $\SO(3,1)$-group, while the $\delta$-distribution of the second line is on the coset.
In particular, we evaluate the convolution at the identity $a=e$, which corresponds to a zero-momentum $\vec{P}=0$~:
\bes\label{coset convolution1}
\phi \dia \psi (e)&=& \int [da_1][da_2]\, \phi(a_1)\psi(a_2)\delta(a_{12})\nn\\
&=& \int [dP]^2\, \phi(P_1)\psi(P_2)\delta(P_1\oplus P_2).
\ees
This convolution product is also non-associative, we need to keep track of the grouping of the convolution products we take. Thanks to this convolution product, we can introduce our proposal for a $\phi^3$ type scalar field action defined on the homogenous momentum space $\hh_3$.
\bes\label{action field dia}
S_\dia(\phi)&=& \int [da]^2\, \phi(a_1)(\vec P_1^2(a)-m^2)\phi(a_2)\int[dh]\,\delta(a_1a_2h)+ \frac{\lambda}{3!} \phi\dia(\phi\dia\phi) (e)\nn\\
&=& \int [dP]^2\, \phi(P_1)(\vec P_1^2-m^2)\phi(P_2)\,\delta(P_1\oplus P_2) \nn\\
&&+ \frac{\lambda}{3!} \int [dP]^3\,\phi(P_1) \phi(P_2)\phi (P_3)\,\delta(P_1\oplus(P_2\oplus P_3)).
\ees

\section{Fourier transform and Snyder space-time}
We now construct  a generalized Fourier in order to define the dual space-time. We first introduce the plane-wave  $e^{iP\cdot x}=e^{iP(a)\cdot x}$ with $x_\mu\in \R^3$ and star product noted $\star$ between the plane-waves in order to represent the modified momenta addition  \eqref{non assoc add}~:
\bes \label{product planewave}
e^{iP_1\cdot x }\star e^{iP_2\cdot x }&\equiv& e^{i(P_1\oplus P_2)\cdot x }
\ees
We define the Fourier transform of a distribution $\phi$ as
\be
\hat \phi(x)\equiv \int [da] \,  e^{iP(a)\cdot x  } \phi(P) ,
\ee
The $\star$ product is the dual of the convolution product
\bes
\int [da]\,  e^{iP(a)\cdot x}  (\psi\dia \phi)(a) &=&\int [da_i]^2\, \psi(a_1)\phi(a_2)\int [da][dh]\,   e^{iP(a)\cdot x}  \delta(a\mone a_1a_2 h )\nn\\
&=&  \hat \psi \star \hat \phi (x)
\ees
Since the convolution product is non-associative, the $\star$ product will also be non-associative. We explore the properties of this $\star$ product by considering the products between monomials \cite{snyder gili}.
\bes\label{1st order monomial}
x_\mu \star x_\nu& =&x_\mu\,x_\nu \\
x_\mu \star (x_\nu\star x_\alpha)&=& -\frac{1}{\ka^2}\delta_{\nu\alpha}x_\mu +x_\mu x_\nu x_\alpha\neq ( x_\mu \star x_\nu) \star x_\alpha. \label{2nd order monomial}
\ees
The $\star$ product encodes therefore some type of non-commutativity  in a slightly different way than usual: the $\star$ product between two coordinate functions is commutative but the position operators still do not commute. To understand the non-commutative geometry structure behind this construction we introduce the operator positions which act by $\star$-multiplication by $x_\mu$, and calculate  the commutator  $[X_\mu,X_\nu]$ for example on the function $x_\alpha$ using \eqref{2nd order monomial}:
\bes
[X_\mu,X_\nu]\act x_\alpha&=& x_\mu\star (x_\nu\star x_\alpha)-x_\nu\star (x_\mu\star x_\alpha)\nn\\
&=& \frac{1}{\ka^2}\left(\delta_{\mu\alpha}x_\nu -\delta_{\nu\alpha}x_\mu\right)= i\frac{1}{\ka^2}J_{\mu\nu}\act x_\alpha,
\ees
where $J_\mn\in \so(3)$ act in the usual way on the space-time. Using the Fourier transform, this calculation can be extended to an arbitrary function $f$ and we see therefore that
\be \label{snyder bracket}
[X_\mu,X_\nu]=  i\frac{1}{\ka^2}J_{\mu\nu}.
\ee
This commutator properly encodes the commutation relation of the Snyder coordinates \cite{snyder}. \emph{We have therefore constructed  a realization of the (Euclidian) Snyder space-time in terms of a star product}.

The commutator of the \emph{classical} coordinates is zero following \eqref{1st order monomial}.
$$[x_\mu,x_\nu]_\star= x_\mu\star x_\nu- x_\nu\star x_\mu=0$$
This might seem puzzling if one has in mind the commutator \eqref{snyder bracket}. This however should not come as a surprise. Indeed when working with the star product, we work at the level of the classical (deformed) algebra of continuous functions $\cc_\star$ over Minkowski space. The star product is therefore stable in $\cc_\star$ and it is not possible to have a commutator of the type $[x_\mu,x_\nu]_\star\sim J_\mn$ since $J_\mn$ is not in $\cc_\star$. However at the level of \emph{operators}, both $X_\mu\sim J_{0\mu}$ and $J_\mn$ have a well defined action on $\cc_\star$, so that the commutator $[X_\mu,X_\nu]\sim J_\mn$ does make sense. To our knowledge there is no other example of non-commutative space where the commutator $[x_\mu,x_\nu]_\star$ is actually different than its operator representation $[X_\mu,X_\nu]$ where $X_\mu$ acts by $\star$-multiplication. The difference can be traced back to the non-associative structure inherent to the coset.

\section{Symmetries}
By construction the action of $H=\SO(3)$ on the coset is given by the adjoint action, which in terms of the Snyder coordinates, is simply the usual "Lorentz" action $$a\dr hah\mone \,\Rightarrow\,  [J_{ij},P_l]=\delta_{jl}P_i-\delta_{il}P_j, \quad J_{ij}\in \su(2).$$ The measure $[da]= d^3P\, \left(1- \frac{\vec P^2}{\kappa^2} \right)\mone$  is clearly invariant under the adjoint action of $H$. The distributions  $\phi,\delta$ are transformed in the standard way
$$\phi(a)\,\dr\, \phi(hah\mone), \qquad\delta(a)\,\dr\, \delta(hah\mone). $$
We note in particular that a function $f$ on the coset evaluated at the origin $e$ is then left invariant under the adjoint action of $H$. As a direct consequence, the convolution product $\phi\dia\psi(e)$  will be invariant as well. With these properties in mind, it is straightforward to check that the action \eqref{action field dia} is invariant under the adjoint action of $\SO(3)$ (\ie the "Lorentz transformations").

To introduce the notion of translations, we use space-time. We shall then realize these transformations in momentum space in order to check that the action has also some kind of translation invariance.
 \bes
\hphi(x+\epsilon)&=& \int [dP]\, \phi(P)  e^{iP\cdot(x+\epsilon) }\\
(\hphi \star \hphi)(x+\epsilon)&=& \int [dP_i]^2\, e^{i(P_1\oplus P_2)\cdot (x+\epsilon) } \phi(P_1)\phi(P_2)  \nn
\ees
We see therefore that in momentum space the translations act a phase multiplication. Moreover, when dealing with many  fields, we  use the $\star$ product between the plane-waves to define  the transformation of the star product of fields.
 \bes
 \phi(P)&\dr& e^{iP(a)\cdot \epsilon} \phi(P)\\
\phi(P_1)\phi(P_2)& \dr&  e^{iP_{1}\cdot\epsilon} \star e^{iP_{2}\cdot\epsilon} \phi(P_1)\phi(P_2) =e^{i(P_{1}\oplus P_2)\cdot\epsilon} \phi(P_1)\phi(P_2) .
\ees
Thanks to the Dirac delta function, encoding the conservation of momentum, we see therefore that the action is invariant under the translations as well. For example we have
\bes
\int [dP]^2\,\phi(P_1) \phi(P_2)\,e^{i(P_1\oplus P_2)\cdot \epsilon}\delta(P_1\oplus P_2)= \int [dP]^2\,\phi(P_1) \phi(P_2)\,\delta(P_1\oplus P_2).
\ees
The action \eqref{action field dia} is therefore invariant under a deformation of the Poincar\'e group. The Lorentz part is not deformed, indeed Snyder noticed already that the non-commutative structure is consistent with the Lorentz symmetries \cite{snyder}. On the other hand  the translations sector is deformed in a  consistent way  with the non-commutative structure. We do not expect this type of deformation to be among the ones  classified in \cite{zak}. Indeed, this deformation is non-(co)associative whereas the classification \cite{zak} looked at the deformations which preserved the (co)associativity. It would be interesting to see if the quantum group we are dealing with could be encoded in a type of quasi-Hopf algebra, that is if there is a 3-cocycle that would rule the lack of (co)associativity (cf for example \cite{majid}).

\section{Concluding remarks}
We have constructed an action for a scalar field with momentum space a coset. We used the properties of this coset to define a modified momenta addition. We then introduced a generalized Fourier transform which allowed to defined the dual non-commutative space-time, which show to be a realization of Snyder space-time. The key feature of the construction is the non-associativity of the different products we used, property which is traced back to the geometry of the coset.

A non-associative addition of momenta is something that is definitely physically difficult to understand from the usual field theory perspective. One can therefore try instead to introduce an associative convolution product on the coset. Using  the group product from $G$   we define then a convolution product based on the group product on $G$. Starting with $H$-invariant fields on $G$, $\phi(g)=\phi(gh),\,\forall h\in H$, we can define their usual convolution product:
$$
\phi \circ\psi (g)=\int [dg_1][dg_2]\,  \phi(g_1)\psi(g_2)\,\delta( g\mone g_1g_2 ).
$$
Since the resulting convolution is still $H$-invariant, this can be written as a convolution product on the coset $G/H$ directly~\footnotemark:
\bes\label{G convolution}
\phi \circ\psi (a)&=& \int [da_1][da_2]\,  \phi(a_1)\psi(a_2)\int [dh_1][dh_2]\,\delta( a\mone a_1 h_1 a_2 h_2  ) \nn\\
&=&  \int [da_1][da_2]\,  \phi(a_1)\psi(a_2)\int [dh_1]\,\delta(a\mone\cdot  a_1 \cdot (h_1 a_2 h_1\mone)  ).
\ees
\footnotetext{We could start with the equivalent definition $\phi \circ\psi (a)= \int [da_1][da_2]\,  \phi(a_1)\psi(a_2)\int [dh_1][dh_2][dh]\,\delta( a_1 h_1 a_2 h_2 h\mone a\mone)$ and use the right invariance of $[dh_2]$ to eliminate $h$. Nevertheless, this definition is divergent if $H$ is non-compact. We therefore work with \eqref{G convolution}.}
Expressed in term of the Snyder momentum, we get:
\be
\phi \circ\psi (P)\,=\,
\int [dP]^2\,  \phi(P_1)\psi(P_2)\int [dh]\,\delta((-P)\oplus (P_1\oplus h\rhd P_2)).
\ee
First, we see that the integral over $h$ averages over the direction of the second momentum $\vec{P}_2$, thus the convolution loses all information about this direction and only remembers the modulus of $\vec{P}_2$. In other words, $\phi \circ\psi$ does not depend on the entire field $\psi(P)$ but only on its radial component $\tl{\psi}(P)=\int dh \psi(h\rhd P)$.

Moreover, this convolution product can not be expressed simply in term of a deformed addition of momenta. At the end of the day, we obtain a non-trivial distribution of the resulting (final) momentum $\vec P$ which not only depends on the initial momenta $P_1$ and $P_2$ but also on the arbitrary group rotation $h$. This seems to be the price in order to have an associative convolution product, and thus star-product, on the coset.

For these reasons, this associative product does not seem to be neither physically motivated, nor interesting from the usual quantum field theory point of view. As a conclusion it seems that the natural non-associative convolution product inherent to the coset is the most interesting structure to study from the quantum field theory approach. It provides us with a proper $star$-product representation of the non-commutative Snyder space-time. The next step will be to study the quantization of the (scalar) field theory on the Snyder space-time and to see whether a Fock space representation is possible or not. The present $\star$-product will likely be instrumental in this development.


\begin{theacknowledgments}
F.G. wants to thank the organizers of the conference, in particular J. Kowalski-Glikman, for their kind hospitality.
\end{theacknowledgments}

\bibliographystyle{aipproc}   

\end{document}